\documentclass[twocolumn,aps,prl,nofootinbib,groupedaddress,amsmath,amssymb,longbibliography]{revtex4-1}
\usepackage{graphicx}
\usepackage{color}
\usepackage{lineno}
\usepackage{bm}
\usepackage{float}
\usepackage{changes}
\usepackage{braket}

\newcommand{\nR}{n_{\rm R}}
\newcommand{\gammac}{\gamma_{\rm C}}
\newcommand{\gammar}{\gamma_{\rm R}}
\newcommand{\gr}{g_{\rm R}}
\newcommand{\gc}{g_{\rm C}}

\begin{document}
	
	\title{A topological attractor of vortices as a clock generator based on polariton superfluids}
	\author {Xuemei Sun$^{1}$, Gang Wang$^{1,*}$, Kailin Hou$^{1}$, Huarong Bi$^{1}$, Yan Xue$^{1,\dagger}$ and Alexey Kavokin $^{2,3,\ddagger}$}
	\affiliation{$^{1}$ College of Physics, Jilin University, Changchun 130012, P. R. China}
	\affiliation{$^{2}$Westlake University, School of Science, 18 Shilongshan Road, Hangzhou 310024, Zhejiang Province, China}
	\affiliation{$^{3}$Westlake Institute for Advanced Study, Institute of Natural Sciences, 18 Shilongshan Road, Hangzhou 310024, Zhejiang Province, China}
	\affiliation{email: $^{*}$wg@jlu.edu.cn, $^{\dagger}$xy4610@jlu.edu.cn, $^{\ddagger}a.kavokin@westlake.edu.cn$}
	
	\begin{abstract}
		We reveal a topologically protected persistent oscillatory dynamics of a polariton superfluid, which is driven non-resonantly by a super-Gaussian laser beam in a planar semiconductor microcavity subjected to an external C-shape potential. We find persistent oscillations, characterized by a topological attractor, that are based on the dynamical behavior of small Josephson vortices rotating around the outside edge of the central vortex. The attractor is being formed due to the inverse energy cascade accompanied by the growth of the incompressible kinetic energy. The attractor displays a remarkable stability towards perturbations and it may be tuned by the pump laser intensity to two distinct frequency ranges: 20.16$\pm$0.14 GHz and 48.4$\pm$1.2 GHz. This attractor is bistable due to the chirality of the vortex. The switching between two stable states is achieved by altering the pump power or by adding an extra incoherent Gaussian pump beam.
	\end{abstract}
	
	\maketitle
	\section{Introduction}
	Exciton-polariton condensates \cite{2006nature_Dang,2009natphy_Bramati,Yamamoto2010} offer intriguing possibilities for the laboratory simulation of a broad variety of non-linear phenomena in classical \cite{2009prl_Shelykh,2019natphon_Lagoudakis,Berloff2017} and quantum regimes \cite{2010prl_cuiti,2019nc_Deveaud,Liew2021}. 
	The nonlinearity arises from the fact that the material component of exciton-polaritons allows them to interact between themselves \cite{Kavokin2007,Amo2017}. On top of this, being bosonic quasiparticles, polaritons possess the remarkable ability to form superfluids even at high temperatures \cite{2014pnas_Dreismann,Forrest2010,Xiong2018}. This superfluid behaviour is conveniently described by a many-body wavefunction that is governed by a generalized Gross-Pitaevskii equation. The generalization is required to account for pumping and dissipation that are always present in any polariton system because of the finite (and usually very short) lifetime of each individual polariton. The driven-dissipative nature of polariton superfluids is responsible for a multitude of fascinating phenomena \cite{2018Suchomel,2020science_Songqh,2021_Dikopoltsev,2020prx_Carusotto}.	To be more specific, thanks to pumping and dissipation, every polariton system is out of thermal equilibrium and it doesn't strive to minimize its energy. Polariton condensates may be formed in an excited state of the lower polariton dispersion branch \cite{Toikka2020,Wouter2019,Kavokin2023} or even in a superposition of excited states \cite{Lagoudakis2020,Demirchyan2014,Tim2016} of the conservative Hamiltonian of the system as long as the pumping is on. 
	
	The stationary state of a superfluid is governed by the balance of pump and decay. If the initial state of the superfluid is a superposition of two eigen-states of the hermitian part of the Hamiltonian of the system, and this state is close to the stationary state, the subsequent dynamics of the superfluid may be characterized by an exciting phenomenon of persisting oscillations. Such oscillations manifest themselves through periodic modulations of the density of polaritons, the phase of the condensate, or through the spatial redistribution of either of the two. In this latter case, oscillations of quantized vortices\cite{Roumpos2011,2019prl_Ostrovskaya} may be found. Quantized vortices are subject to a detailed experimental study in polaritonics. They can be conveniently identified by near-field interferometry.
	
	A variety of physical mechanisms may be behind the formation of persistent currents or vortices in polariton superfluids. A comparison of the polariton system to ac-Josephson currents of Cooper pairs might be instructive or deceptive, depending on how the system responds to its initial conditions. If the  initial state of the system is formed by a superposition  of two eigen-states of the hermitian component of the Hamiltonian, the ac-Josephson current is formed \cite{xue2021,Palavous2023}. In contrast, if the system's stationary dynamics and the initial state are mismatched, a chaotic dynamics might emerge to provoke an inverse energy cascade\cite{Ecke2012,Sanvitto2023} where vortical structures from small to large ones form increasingly over time. The appearance of co-rotating pairs of vortices \cite{Haziot2023} is one of the signatures of this mechanism \cite{Takaoka1994,Ashida2022}.
	
	In this work, we report a persistent nonlinear dynamics of a vortex attractor in a circular polariton superfluid, where the central large vortex can capture Josephson vortices being relatively small objects that consistently circulate around it. By analyzing the dynamics of the system, we establish that the appearance of co-rotating pairs of vortices, induced by the inverse energy cascade intrigued during the dynamical procedure of the chaotic phase of repeated collisions of vortices, is crucial to the formation of this topological attractor. Topological protection ensures that the dynamics of attractors, which display persistent oscillations throughout time, are unaffected by external perturbations. The system may be seen as a "clock generator" \cite{2020prb_Solnyshkov} which presents a remarkable stability and may be adjusted to two distinct frequency domains by tuning the pump laser intensity. Since a vortex has a chiral symmetry, every attractor is bistable. In each particular numerical experiment, the system spontaneously relaxes to one of two stationary solutions. These findings provide a proof-of-concept demonstration of the feasibility of a portable and low-power clock generator based on a polariton superfluid.
	
	\section{Model} 
\begin{figure} [!h]
		\centering
		\includegraphics[angle=0,width=9.0cm]{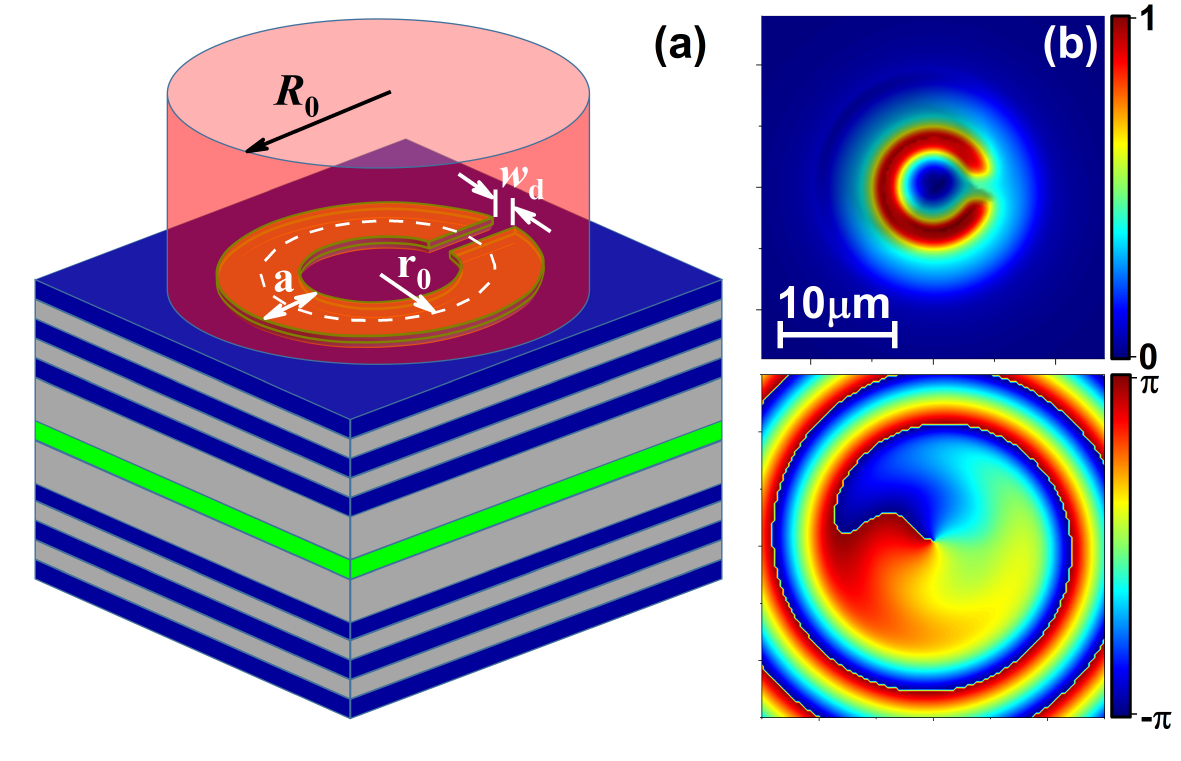}
		\caption{ (a)Sketch of the semiconductor microcavity containing a C-shaped potential, excited by a non-resonant pump having the super-Gaussian shape with the radius $R_{0}=10\mu$m. (b) stable state of vortex achieved with $P_{0}=3.1P_{th}$. The parameters for a C-shaped potential with a slot are: $r_{0}=4\mu$m, $a=2\mu$m, $w_{d}=2\mu$m and the potential depth $V_{0}=-0.6$meV. }
		\label{fig:sketch}
	\end{figure}

	To be specific, we consider a system shown schematically in Fig. \ref{fig:sketch}(a). The non-resonant CW optical field of a super-Gaussian shape $P(r)=P_{0} \times e^{-(\frac{r}{R_{0}})^{20}}$ is applied to excite exciton-polaritons in a semiconductor microcavity containing a C-shape in-plane external potential. $r_0$ and $a$ are the radius and width of the potential ring, respectively. $V_{0}$ is the potential depth and $w_d$ is the width of the potential slot. Obeying the bosonic statistics, polaritons form a Bose-Einstein condensate which remains localized under the joint confinement effect of the external potential and the pump spot.
	
	The dynamics of a polariton condensate can be described by the dissipative Gross-Pitaevskii (GP) equation for the macroscopic wave function $\psi(\textbf{r},t)$, coupled to the rate equation for the density of the exciton reservoir $\nR(\textbf{r},t)$:
	
	\begin{align}\label{mean}
		i\hbar\frac{\partial \psi(\textbf{r},t)}{\partial t} =
		&[-\frac{\hbar^2}{2 m}\nabla^2  +  \gc |\psi(\textbf{r},t)|^2 + \gr \nR(\textbf{r},t) + V(r) \nonumber\\
		&  + \frac{i\hbar}{2}(R \nR(\textbf{r},t)- \gammac)] \psi(\textbf{r},t) +i\hbar \frac{dW}{dt} \nonumber \nonumber \\
		\frac{\partial \nR(\textbf{r},t)}{\partial t}=
		&P(\textbf{r})-(\gammar +R |\psi(\textbf{r},t)|^2 )\nR(\textbf{r},t) 
	\end{align}
	here $m=1\times 10^{-4}m_{e}$ (${m_{e}}$: free electron mass) is the effective mass of polaritons on the lower-polariton branch. The nonlinear coefficients $\gc=3 \times 10^{-3} $meV$\mu$m$^2$ and $\gr=2gc$ represent the strengths of polariton interactions between themselves and with the reservoir exciton, respectively. $\gammac=0.4 ps^{-1}$ and $\gammar=0.8 ps^{-1}$ are the polariton and the reservoir decay rates, respectively. $R=0.01$ps$^{-1}\mu$m$^2$ is the rate of stimulated scattering of quasiparticles from the exciton reservoir to the polariton fluid. $P(\textbf{r})$ is the non-resonant cw optical pump.	$V(\textbf{r})$ is the external potential with depth that can be generated in planar semiconductor microcavities by different techniques \cite{2007Science_West,2007nature_Yamamoto}. $dW$ describes the quantum fluctuations within the classical field approximation by the addition of a complex stochastic term in the truncated Wigner approximation (TWA)
	\begin{align} \label{stochas}
		<dW(\textbf{r},t)dW(\textbf{r}',t')>&=0  \\
		<dW(\textbf{r},t)dW^*(\textbf{r}',t')>&=\frac{dt}{2 dxdy}(R \nR+ \gammac) \delta_{\textbf{r},\textbf{r}'} \delta_{t,t'} \nonumber
	\end{align}
	
	\section{Topological attractor} 
\begin{figure} [!htbp]
		\centering
		\includegraphics[angle=0,width=9.0cm]{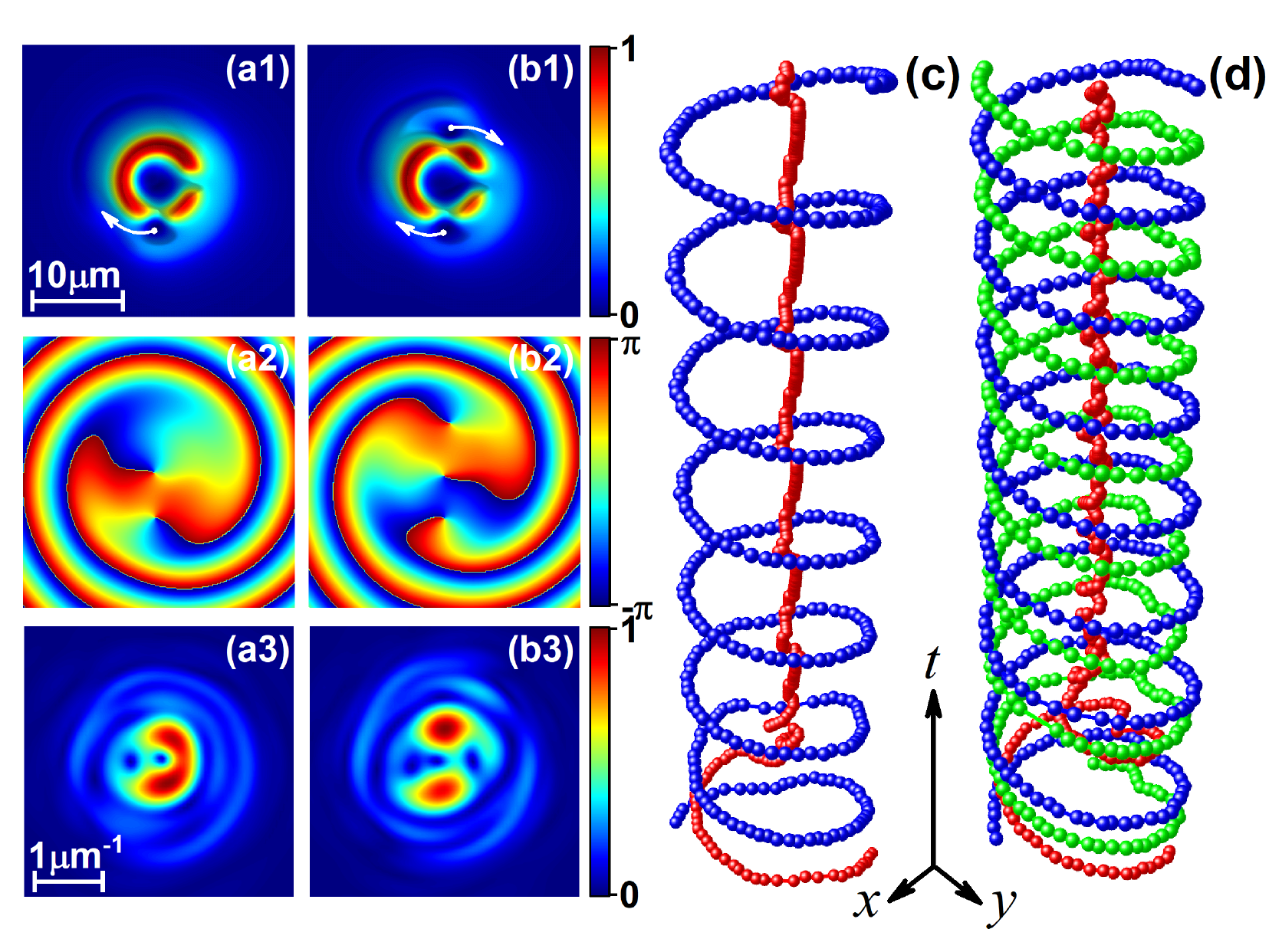}
		\caption{ Vortex attractor. (a1-b1) polariton density in real space, (a2-b2) phase distribution in real space and (a3-b3) polariton density in momentum space for two different pump power. The central vortex captures (a) one small Josephson vortex ($P_{0}=2.9P_{th}$) or (b) two small Josephson vortices to circle around it ($P_{0}=2.6P_{th}$). White arrows depict the direction of movement of the Josephson vortices. (c-d) the trajectory of the vortex singularity in the $xyt$ space for the vortex attractor in (a) and (b).}
		\label{fig:attractor}
	\end{figure}

	The polariton superfluid, due to its non-hermitian nature, requires persistent external pump to sustain the polariton population by means of the stimulated scattering of quasiparticles from the exciton reservoir to the polariton reservoir. The spatial characteristics of the pump have a notable influence on the reservoir-induced effective potential confining the condensate, which, in conjunction with the external potential, plays a crucial role in determining the dynamics of the superfluid system. For instance, in the case of a polariton superfluid confined within an external C-shaped potential, although Josephson vortex pairs are formed at the location of potential slot functioning as a Josephson junction, the ultimate stationary state hinges on the profiles of the pumping mechanism. If the pump exhibits an annular configuration, the superposition of two eigen states of the hermitian part of the Hamiltonian of the system can be excited and the subsequent dynamics of the superfluid may be characterized by quantum beats with a characteristic frequency dependent on the energy splitting of the involved eigen states \cite{xue2021}. If the pump, however, exhibits a super-Gaussian profile, only one excited state with two modes is stimulated and the subsequent dynamics of the superfluid may manifest themselves in periodical oscillation where a central vortex captures Josephson vortices to orbit around it. We shall refer to this spectacular phenomenon, demonstrated in Fig. \ref{fig:attractor}, as to a vortex attractor.
	
	Two types of topological attractor, presented in Fig. \ref{fig:attractor}, emerge spontaneously due to the random fluctuations of the initial noise and the quantum fluctuation dW. The topological attractor A, as seen in panels (a1-a3), illustrates the periodical rotation of a single small Josephson vortex around the outer edge of the annular density of the central vortex. Conversely, panels (b1-b3) depict the periodic rotation of two small Josephson vortices, which is labeled as topological attractor B. The former is accomplished by utilizing a pump power of $P_{0}=2.9P_{th}$ while the latter is obtained by employing a slightly lower pump power of $P_{0}=2.6P_{th}$. In both scenarios, all small Josephson vortices exhibit identical circulation patterns to that of the central vortex, as depicted in the phase diagram. It is clear that the quantity of the Josephson vortices, which are captured as small objects by the central vortex, diminishes as the pump power increases.  Hence, it is not surprising that a stable vortex state can be observed as the pump power is progressively raised to the value of $P_{0}=3.1P_{th}$, as seen in Fig. \ref{fig:sketch}(b). This observation is consistent with the non-hermitian nature of the polariton superfluid, wherein an increase in the pump power leads to the decrease in the energy of the state. The effects revealed by this study diverge from those observed in polariton condensates subjected to annular external potentials. In the context of the aforementioned study \cite{2020ol_ma}, it is observed that vortex structures characterized by varying winding numbers (representing angular momentum) tend to exhibit stability when subjected to adjustments in the pump power. The limit cycles have been discussed and experimentally searched for in polariton fluids. Proposals for polariton time crystals \cite{Gumbs2023,Shelykh2019} based on such limit cycles have been made. In this context, the originality of the present work is in the focus on vortex/antivortex oscillation dynamics that allows for a relatively easy experimental detection via interferometry and brings an interesting phenomenology of the Laguerre-Gaussian beams with oscillation orbital momenta.
	
	It is important to highlight that in the context of the topological attractor, both the central vortex and the small Josephson vortices are close to the first excited state of the external potential where a stable vortex is normally located. The polariton density distribution in momentum space, as seen in Figs. \ref{fig:attractor} (a3, b3), illustrates the presence of small Josephson vortices within the annulus of the density of the central vortex. These small Josephson vortices split the central vortex into several portions to generate unbalanced modes in the first excited state. The binding energy between the central vortex and the small Josephson vortices may be approximated by calculating the vector difference of their singularity locations in momentum space. It is worth noting that the uncertainty of this estimation, denoted as $\Delta k$ is about 0.08$\mu m^{-1}$. 
	
	Figs. \ref{fig:attractor}(c-d) illustrates the spatio-temporal trajectories of the vortex singularities within the studied topological attractor, with a temporal resolution of $\delta t=1$ps. The singularity of the central vortex remains intact as it moves within the core area, but the singularities of the rotating Josephson vortices follow a circular course. The orbital-like trajectory is analogous to that of the half-vortex in a spinor polariton condensate \cite{2015_spinor-half-vortex_sanvitto}, where a vortex characterised by a certain circular polarization coincides with a Gaussian beam possessing the opposite circular polarization. Orbital-like trajectories suggest the attractive interaction between the central vortex and the rotating Josephson vortices.
	
	\begin{figure*} [!htbp]
		\centering
		\includegraphics[angle=0,width=16.0cm]{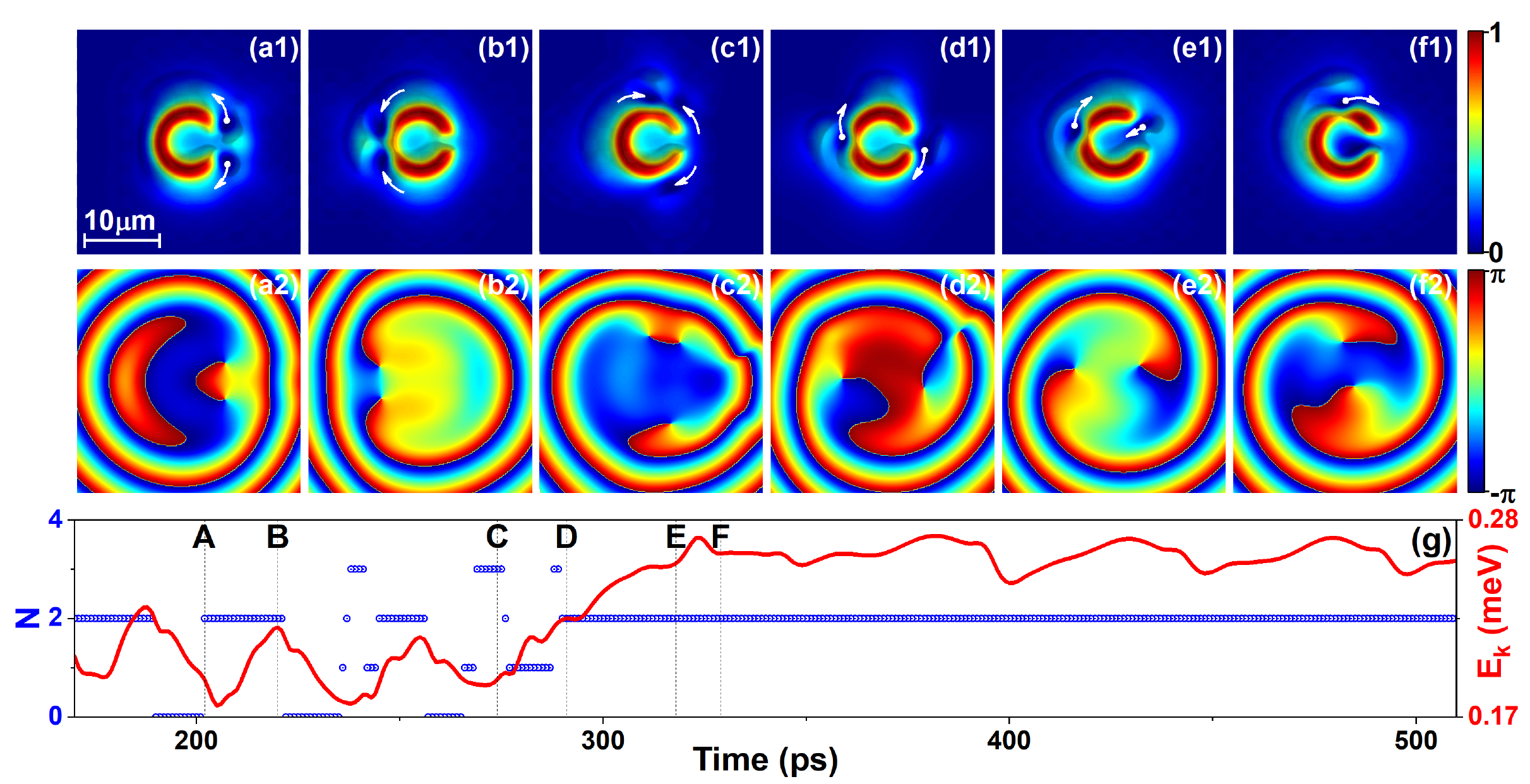}
		\caption{ (a1-f1) The polariton densities and (a2-f2) the corresponding phases in the dynamical procedure for the formation of a topological attractor having a single Josephson vortex orbiting around the central vortex. (g) the winding numbers and kinetic energy vs time. The inverse energy cascade is primarily responsible for the formation of an attractor.}
		\label{fig:procedure}
	\end{figure*}
	
	In order to reveal the physical mechanism responsible for the formation of a topological attractor, it is essential to recall the physics of an inverse energy cascade. This cascade involves the increase of the incompressible kinetic energy per vortex, leading to the development of large vortical structures over time. The study conducted by Sanvitto et. al. \cite{Sanvitto2023} has successfully illustrated the occurrence of vortex clustering in a polariton system, highlighting the tendency of the vortex-gas towards highly excited configurations. In the following, we shall concentrate on the role of the inverse energy cascade in generation of a co-rotating pair of vortices characterised by identical winding numbers  (see Fig.~\ref{fig:procedure}(g)), which plays a critical role in the dynamics of a topological attractor. 
	
	Let us consider the scenario involving a single Josephson vortex playing role of a small rotating object. The polariton density generated by the pump with $P_{0}=2.9P_{th}$ is sufficient to ensure the superfluid behaviour of the polariton condensate, resulting in a shallow density distribution across the potential slot. This distribution disrupts the rotational symmetry of the polariton density and exhibits features similar to a Josephson junction. Consequently, Josephson vortex-antivortex pairs are induced symmetrically with respect to the \rm{y}-axis on the right side of the potential slot (see Fig.~\ref{fig:procedure}(a)). The vortex and anti-vortex exhibit opposite rotational orientations around the outside border of the ring-shape dense part of the polariton condensate and intersect at a specific spot symmetric to the potential slot with respect to the \rm{x}-axis (see Fig.~\ref{fig:procedure}(b)). Following the collision, the vortex and anti-vortex tunnel through the dense part of the ring and subsequently undergo annihilation. Although the observed dynamical activity exhibits periodicity throughout several cycles, it does not represent a stationary state of the system. At a certain moment, this behavior is abruptly interrupted, resulting in the cessation of tunneling following the collision. One vortex loses its energy following the impact and progressively diminishes over time, while the other vortex persists in its rotational motion at the outer boundary of the ring-shape dense part of the polariton condensate  (see Fig.~\ref{fig:procedure}(c)). Therefore, the left vortex inevitably hit with the newly generated vortex or antivortex. In a 2D condensate, the occurrence of repeated collisions between vortices can give rise to an inverse energy cascade, which is facilitated by the inclusion of the quantum pressure term ($|\psi|^2\nabla(\frac{\hbar^2}{2 m}\frac{\nabla^2 |\psi|}{|\psi|})$) \cite{Primer_2016} in the Gross-Pitaevskii equation. As a result of this cascade, the vortex reconnection takes place, leading to the annihilation of vortices possessing a particular circulation and the generation of a co-rotating vortex pair with the same circulation.  This phenomenon is shown in Fig.~\ref{fig:procedure}(d): the right-circulation vortex is eliminated following the collision, whereas a co-rotating pair of vortices with left-circulation is retained. 
	
	Fig.~\ref{fig:procedure}(g) illustrates the fluctuations in vortex counts and the associated time dependence of the incompressible kinetic energy. The temporal intervals represented by points A to F correspond to those in the  figures (a) to (f), respectively. It is evident that prior to the formation of a dynamics shown in C, the annihilation of the vortex through tunneling consistently coincides with a reduction in the kinetic energy. On the other hand, it is observed that the kinetic energy between points C and D is steadily growing, despite the decrease in the vortex numbers resulting from collisions. The occurrence of a co-rotating pair of vortices provides a direct proof of the inverse energy cascade.
	
	The events following the formation of a pair of co-rotating vortices are easy to interpret. Under the effect of Magnus force $F_{M}=2\pi \hbar m |\psi|^2 \vec{e}_{z} \times \vec{v}_{rel}$ \cite{Sonin_1979}, the pair of vortices keeps rotating around the outside broader of the ring-shape dense part of the condensate until one of the vortices encounters the potential slot. The shallow polariton density at the potential slot modifies the effective Magnus force to drag the small Josephson vortex into the core of the ring-shape condensate (see Fig.~\ref{fig:procedure}(e)). This spiraling induces the excitation of a central vortex while it does not alter the trajectory of the remaining vortices, which continue to behave as small objects persistently rotating around the central vortex (see Fig.~\ref{fig:procedure}(f)). All together, this produces a dynamically stable topological attractor. Once the direction of the periodic oscillation is established, it would never change. 
	
	\begin{figure} [!htbp]
		\centering
		\includegraphics[angle=0,width=8.0cm]{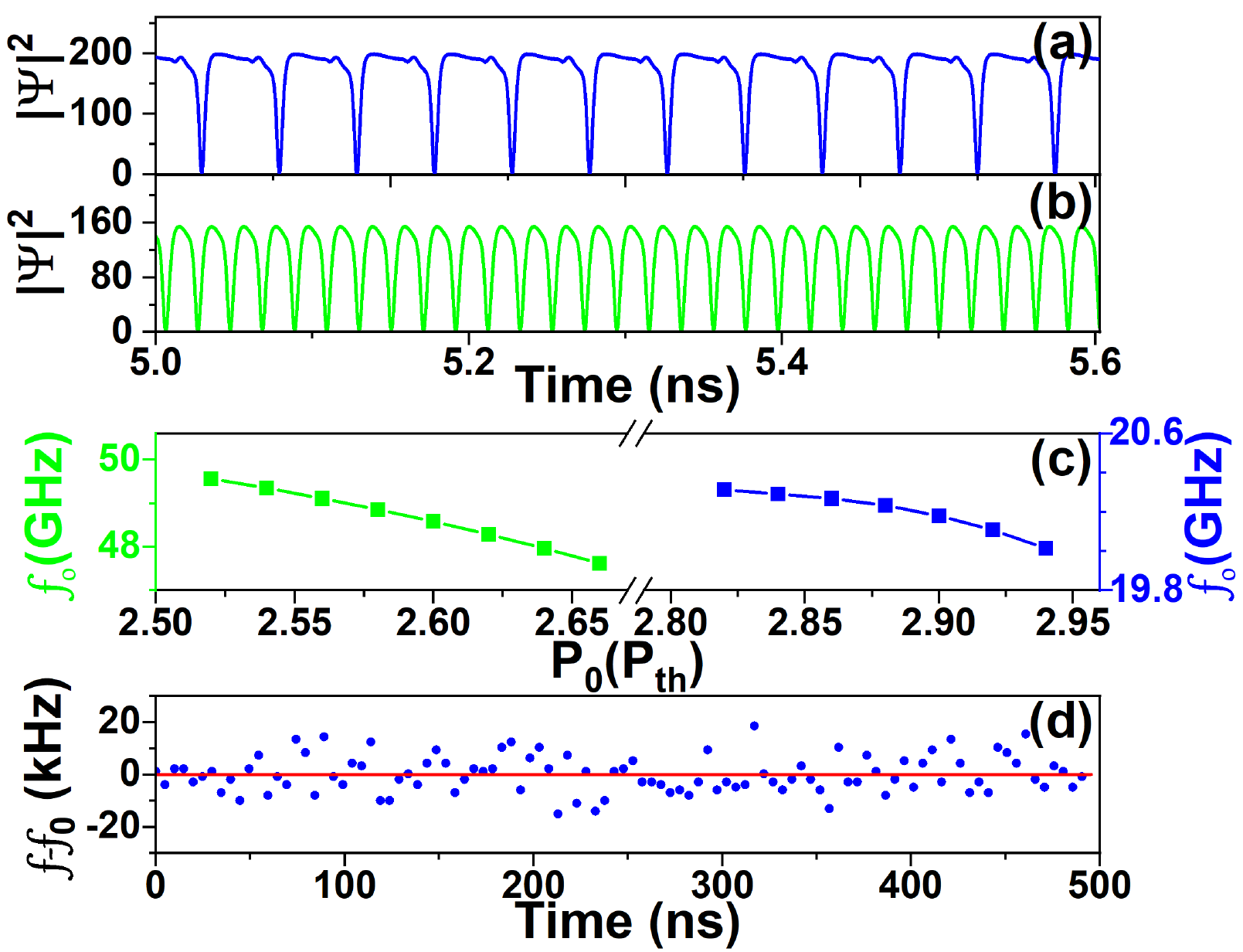}
		\caption{The frequency of oscillations exhibited by the polariton optical clock. (a) and (b) show the time dependencies of the polariton density at the fixed point of ($x=-5, y=0$) corresponding to Figs. \ref{fig:attractor}a and \ref{fig:attractor}b, respectively. (c) shows the dependence of the frequency of oscillations on the pump power. (d) The time-resolved measurement of the oscillation frequency offset from the mean $\sim$20.18GHz output frequency. }
		\label{fig:period}
	\end{figure}
	
	The periodic rotation of Josephson vortices around the central vortex makes the topological attractor behave like a clock generator. A key figure of merit for clocks is the achievable oscillation frequency and stability. To illustrate these characteristics, we plot in Figs. \ref{fig:period}(a) and \ref{fig:period}(b) the temporal dependencies of the polariton density at the fixed point of ($x=-5, y=0$) for topological attractors A and B, respectively, corresponding to Figs. \ref{fig:attractor}(a) and \ref{fig:attractor}(b). The considered point is passed through by the center of moving Josephson vortices from time to time. High-contrast stable periodic oscillations are visible. These oscillations begin at around 100ps and remain stable for at least 500ns, with no sign of decay or dephasing. We have checked that the period of oscillations over this time range remains perfectly stable though the static potential noise is taken into account in the calculation. Fig. \ref{fig:period}(c) predicts the dependence of the oscillation frequency on the pump power: a parameter that can be tuned easily in the experiments. Two frequency ranges around 20.16$\pm$0.14GHz and 48.4$\pm$1.2GHz are visible. The higher the pump power, the lower the oscillation frequency. We believe that tuning the pump power is the easiest way of tuning the operation frequency, although other parameters affecting the external potential can also be modulated. Fig. \ref{fig:period}(d)  presents the time-resolved measurement of the 20.18GHz clock generator output over 500ns, which corresponds to 10000 periods of oscillations shown in Fig. \ref{fig:period}(a). Here each point is a result of averaging over 100 measurements with an observation time of 100 periods of oscillations. The noise is taken into account according to Eq. (\ref{stochas}). The mean value depicted by the red solid line is in agreement with the simulation results obtained neglecting the noise. The absolute frequency shift of our clock is around $\Delta f \sim$ 20KHz and the corresponding fractional frequency stability is $\Delta f/f_{0} \sim 10^{-6}$.  2D numerical simulations become too expensive if carried over much longer times. This prevents us from predicting here the value of the Allan deviation of the stabilized clock signal to estimate the ultimate stability of the chip-scale polariton clock generator. We believe that an experimental measurement of the Allan deviation would be more important than any theoretical prediction.
	
	\begin{figure} [!htb]
		\centering
		\includegraphics[angle=0,width=8.0cm]{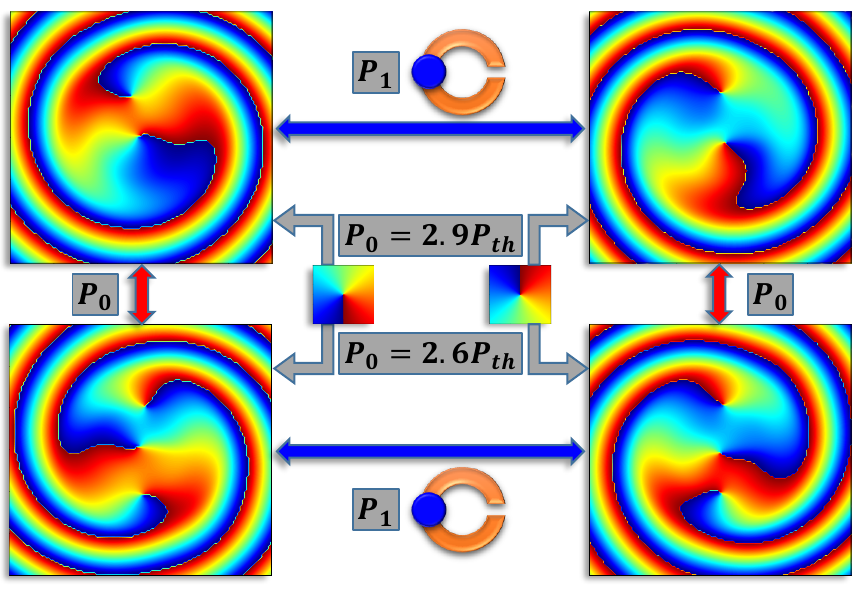}
		\caption{The dynamical switching from any controllable state of the topological attractor to a desired state of topological attractor. The phase profiles of the target topological attractors are shown in the larger panels. The switch between the states with the same directions of rotation is achieved by simply modulating the power of  the incoherent pump field $P_{0}$. The switch between the bistable sates of counter-rotating directions is achieved with an additional incoherent pump $P'(r)$ carrying the same direction of rotation as the target state. Phase profiles of $P'(r)$ and initial seeds are shown in the smaller panels.}
		\label{fig:switch}
	\end{figure}
	
	We also note that the polariton system is bistable due to the strong nonlinearity, which is why it relaxes to either clock-wise or anti-clockwise rotation stochastically, each orientation appearing with a $50\%$ probability. The rotation direction of a given topological attractor may be imprinted by seeding an initial orbital angular momentum, which is similar to the method to control the chirality of stable vortices. Fig. \ref{fig:switch} shows four examples of spontaneously formed topological attractors with target rotation directions defined by seeding an initial orbital angular momentum, which is depicted by two small colored figures. Further, the topological attractor could be switched to a new one characterized by a different frequency by simply modulating the pump power $P_{0}$ (see red arrow). Additionally, it is important to note that the bistable topological attractors with opposite rotation directions could also be controlled by switching an additional incoherent Gaussian pump $P'(r)=P_{1}e^{-[((x+4)^2+y^2)/2^{2}]^{10}}$ at the position symmetric to the potential slot with respect to the \rm{x}-axis (see blue arrows). In general, we are able to switch from any state of a topological attractor to any desired target state. The dynamical switch with the modulation of $P'(r)$ is shown in the supplementary movie, switch-p1-a(b), respectively.
	\begin{figure} [!htb]
		\centering
		\includegraphics[angle=0,width=8.0cm]{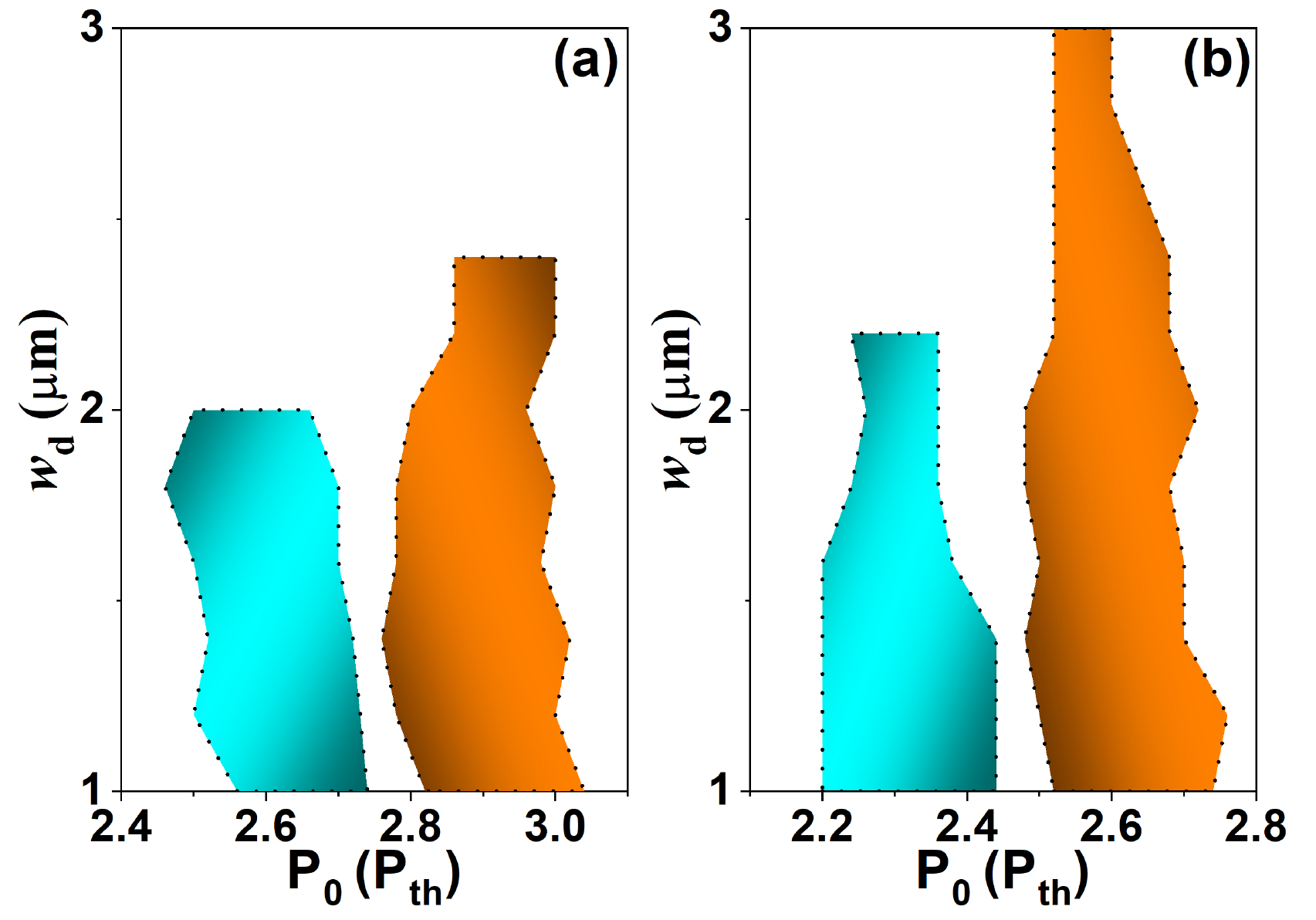}
		\caption{The phase diagram of the existence of the topological attractor shown in Fig. \ref{fig:attractor} for two different potential depth: (a) $V_{0}=-0.6$meV and (b) $-0.5$meV. Orange color for attractor A is shown in Fig. \ref{fig:attractor}(a) and cyan color for attractor B in shown in Fig. \ref{fig:attractor}(b).}
		\label{fig:stability}
	\end{figure}
	
	Through large numerical simulations, we obtain phase diagrams to observe two kinds of topological attractor for two considered potential depths: $V_{0}$ of either $-0.6$meV or $-0.5$meV, shown in Fig. \ref{fig:stability}. To obtain this phase diagram, the dynamics of the system is studied by scanning the pump power $P_{0}$ while keeping a fixed potential slot $w_{d}$. For a fixed $P_{0}$, the topological attractor forms while $w_{d}$ is small enough. In particular, the range of $w_{d}$ where one can observe a topological attractor A characterised by a single Josephson vortex playing the role of a small rotating object is larger than that for a topological attractor B characterised by two Josephson vortices playing the roles of small rotating objects. Comparing two different ranges of the pump power needed to observe the topological attractors for two different values of the potential depth $V_{0}$, one can conclude that the shallower the potential depth, the lower the required pump power.
	
	\section{Conclusion}
	We predict here the formation of topological attractors in a superfluid of exciton-polaritons placed in an external lateral potential of the C-shape geometry. The topological attractor corresponds to the regime where Josephson vortices are rotating periodically around the central large vortex. The inverse energy cascade is shown to be responsible for the formation of topological attractors.	A high stability of the oscillations makes it possible to use the topological attractor as a portable clock generator operating at the low power regime.  We show that the operation frequency of the clock generator can be controlled within two different ranges. Its tuning can be achieved by tuning the laser pump power. 
	
	\section{acknowledges}
	This work was supported by the National Natural Science Foundation of China (Grant No. 12074147). AK acknowledges the support from the Westlake University, Project 041020100118 and Program 2018R01002 funded by the Leading Innovative and Entrepreneur Team Introduction Program of Zhejiang Province.
	

	\end {document}